\documentclass[american,aps,twocolumn]{revtex4}
\usepackage[T1]{fontenc}
\usepackage[latin1]{inputenc}
\usepackage{amsmath}
\usepackage{amssymb}
\usepackage{graphicx}
\makeatletter

\providecommand{\LyX}{L\kern-.1667em\lower.25em\hbox{Y}\kern-.125emX\@}

\usepackage{float}
\usepackage{graphicx}
\usepackage{float}

\usepackage[T1]{fontenc}
\usepackage[latin1]{inputenc}
\makeatletter

\usepackage{babel}
\makeatother

\begin{document}

\title{Electronic structure and optical properties of ZnS/CdS nanoheterostructures}

\author{J. P\'{e}rez-Conde }

\affiliation{Departamento de Física, Universidad Pública de Navarra, E-31006, Pamplona,
Spain }

\author{A. K. Bhattacharjee }

\affiliation{Laboratoire de Physique des Solides, UMR du CNRS, Université Paris-Sud,
F-91405, Orsay, France}

\begin{abstract}
The electronic and optical properties of spherical nanoheterostructures
are studied within the semi-empirical $sp^{3}s^{*}$ tight-binding
model including the spin-orbit interaction. We use a symmetry-based
approach previously applied to
CdSe and CdTe quantum dots. The complete one-particle spectrum is
obtained by using group-theoretical methods. The excitonic eigenstates
are then deduced in the configuration-interaction approach by fully
taking into account the Coulomb direct and exchange interactions.
Here we focus on ZnS/CdS, ZnS/CdS/ZnS and CdS/ZnS nanocrystals with
particular emphasis
on recently reported experimental data. The degree of carrier localization in 
the CdS well layer is analyzed as a function of its thickness. We compute
the excitonic fine structure, i.e., the relative intensities of low-energy
optical transitions. The calculated values of the absorption gap show
a good agreement with the experimental ones. Enhanced resonant 
photoluminescence Stokes shifts are predicted. 
\end{abstract}


\maketitle

%\section{Introduction}

The quantum confinement effects in semiconductor nanocrystals (NC's)
or quantum dots (QD's) have been investigated over the last two decades.
A recent development concerns nanoheterostructures containing concentric layers
of two different semiconductors. The initial idea of capping a NC with a shell
of higher-gap (barrier) material in order to reduce the uncontrolled
surface effects has been
extended to quantum-dot quantum-well (QDQW) structures with an inner
layer of lower-gap (well) material which allow a greater control
over the optical properties.\cite{mk96} There are several examples such as
the inclusion of a layer of HgS in a CdS NC \cite{sm94,km99,mk96},
where the lattice parameters of both compounds are of similar size
($a=5.851$~\AA{} for the $\beta $-HgS and $a=5.818$\AA{} for
the CdS). Recently, ZnS/CdS based QDQW's
have been synthesized \cite{ra98,le01} with a larger lattice mismatch
(more than $7$ \% , the ZnS lattice constant being  $a=5.409$). Some
other combinations have also been studied, such as CdSe/ZnS \cite{dr97}
or CdSe/CdS.\cite{hs99}

On the theoretical side, effective mass approximation
(EMA) models were  proposed \cite{sm94,jb98} without a full
many-body treatment of the Coulomb direct and exchange interactions.
An atomistic theory was needed, however, to adequately describe a
QDQW where the well can be as thin as one monolayer. 
Very recently we proposed a tight-binding (TB) model \cite{pb02}, 
which is a generalization of our previous
work. This theory successfully accounted for the optical properties
of CdSe \cite{pb01} and CdTe \cite{pb99,pb01b} NC's. Independently, Bryant 
and collaborators have also developed a TB approach. \cite{le01,bj01} 
In contrast
with the Bryant group, we include the spin-orbit interaction,
take into account the symmetries
and obtain the full single-particle spectrum
before deducing the exciton states. The Coulomb
and exchange interactions were already incorporated in our study of CdS/HgS/CdS
QDQW's\cite{pb02}, prior to the more recent work by Xie {\em et al.}.\cite{xb02}
To our knowledge, the present paper reports the first TB calculation of the
low-energy excitonic
states in CdS/ZnS nanoheterostructures. After
a brief description of the theoretical model which can be
completed with our previous publications \cite{pb99,pb01,pb02}, we present
our main results on ZnS/CdS, ZnS/CdS/ZnS
and CdS/ZnS nanocrystals, including a comparison with
the available experimental data.

%\section{Theory}

\label{sec:Theory} Our calculations are based on the semi-empirical TB model
for bulk semiconductors
introduced by Vogl \emph{et al.} \cite{vh83} and generalized
by Kobayashi \emph{et al.} to account for the spin-orbit interaction.\cite{ks82}
In this model five atomic orbitals are used: $s$, $p_{x}$, $p_{y}$,
$p_{z}$, $s^{*}$. The phenomenological $s^{*}$ orbital was introduced
in order to simulate the effects of $d$ orbitals on the conduction
bands. The interatomic hopping matrix elements are restricted to the
nearest neighbors. The model
is then described by a total of $15$
independent parameters. The CdS TB parameters used here are essentially
those proposed earlier by Lippens and Lannoo \cite{ll89} which we
have slightly modified in order to incorporate the spin-orbit coupling.
The ZnS parameters are those calculated by Bertho.\cite{be97}
As for the band offset, following Ref. \cite{le01},
the valence band maximum of ZnS is set to $0$ eV and that of CdS 
shifted to $+0.4$ eV. The bulk gaps of ZnS and CdS are $3.7$
eV and $2.5$ eV respectively. 

We outline the method for deriving the symmetrized TB Hamiltonian
(see Ref. \onlinecite{pb99} for details). The NC's, of roughly spherical
shape, are constructed starting from a cation at the origin by successively
adding nearest-neighbor atoms through tetrahedral bonding. We passivate
the NC surface by placing a hydrogen $s$ orbital at each empty nearest-neighbor
site on the surface dangling bond so that the final surface states are
several eV far from the gap edges. 
Hereafter we use a simplified notation: 
We write $r_{co}/r_{w}/r_{cl}$ to indicate that a NC is built with
a core radius $r_{co}$, with $r_{w}$ and $r_{cl}$ designating the
widths of the well and clad layers, respectively. The lengths are all given
in angstroms.

The single-particle TB Hamiltonian is reduced to a block diagonal
form by writing it in a symmetrized basis corresponding to the double-valued
representations $\Gamma _{k}$ ($k=6,7,8$) of the $T_{d}$ point
group.\cite{kd63} This method not only reduces the size of the
Hamiltonian to diagonalize, but also provides
the exact symmetry classification of the wavefunctions. One can thus 
deduce selection rules and related insights over the relative
intensities of optical transitions between the
valence and conduction states.\cite{pb99}
We then diagonalize the Hamiltonian and obtain the full one-particle spectrum
in a given nanostructure:
all the energy levels and eigenstates. We have also calculated the
spatial projections of the density of states (DOS) onto the core, well and
clad regions, when required, to study the influence of the position and 
thickness of different 
layers on the whole energy spectrum.

The optical properties concern exciton-like elementary excitations
which can be described in terms of the Coulomb direct and exchange
interactions between the electron and hole. The total Hamiltonian can
be written as, \cite{pb01}\begin{equation}
H_{vc,v^{\prime }c^{\prime }}=(\varepsilon _{c}-\varepsilon _{v})\delta _{vv^{\prime }}\delta _{cc^{\prime }}-J_{vc,v^{\prime }c^{\prime }}+K_{vc,v^{\prime }c^{\prime }},\label{htotal}\end{equation}
 where $\varepsilon _{c}$, $-\varepsilon _{v}$ are the electron
and hole energies, $J$ and $K$ are the Coulomb and exchange interactions
respectively (see Ref. \onlinecite{pb01} for more details). Note
that we leave the $e-h$ exchange interaction ($K$)
unscreened up to the nearest-neighbor site (primitive cell of the
zincblende crystal), in analogy with the unscreened short-range part
in the theory of bulk exciton.

 The unscreened
on-site Coulomb and exchange integrals for anion (cation) are assumed
to be $U_{coul}=20 (6.5)$ eV and $U_{exch}=1 (0.5)$ eV respectively.
These values follow roughly those obtained for CdSe and CdTe. \cite{lp98,pb01}
The values for the S atom in CdS have been calculated by means of
a simple scaling. We assume the same parameters for ZnS. The nearest-neighbor
exchange integrals are assumed one tenth of the on-site ones. The
permittivity  is taken as $\epsilon (\infty )$:
$\epsilon (\infty )=5.7$ for ZnS and $\epsilon (\infty )=5.2$ for
CdS in single binary QD's. We take an average value in ZnS/CdS, $\epsilon (\infty )=5.4$.
Strictly speaking, the dielectric constant in
a nanocrystal is size-dependent and substantially reduced with respect to
the bulk low-frequency value $\epsilon (0)$. 
See, for example, Ref. \cite{wz96}, where  a pseudopotential calculation
for CdSe nanocrystals is presented:
The dielectric constant increases with increasing size and asymptotically
approaches the bulk high-frequency value $\epsilon (\infty )$. To our knowledge,
no such calculations are available for the present nanoheterostructures nor even
for their components. We, therefore, assume an average of the bulk
high-frequency values. 

The large lattice mismatch in the ZnS/CdS system is expected to lead to elastically strained structures
 with the relatively thin shell/well layers conforming to the substrate lattice constant.\cite{le01} 
 Unfortunately, there is no simple way of including this effect in our TB model. Preliminary calculations 
 suggest that a simple modification of the interatomic matrix elements alone, following the Harrison 
 scaling rule for the bond length, is inadequate: It yields a systematic increase of the optical gap in 
 the ZnS/CdS and ZnS/CdS/ZnS structures considered here by 10-30 \%. The effect is, of course, qualitatively
 inverse in CdS/ZnS. A more refined modelization scheme for structural distortion involving the diagonal 
 matrix elements (see, for example, Ref.\cite{bk02}) seems necessary.
 Work is in progress in this direction and will be reported elsewhere.
 The results presented below
 have been obtained by neglecting the strain effects.

%\section{Results and Discussion}

\label{sec:Results} In Fig. \ref{Density-of-States:a} we show the
total density of states for some ZnS, CdS and ZnS/CdS/ZnS NC's. It
can be seen how the projection of DOS onto the CdS well quickly grows
when one and two monolayers are added. A monolayer yields a CdS well projection,
almost equivalent in size and energy distribution, to the core ZnS
projection. The addition of a second layer leads to a CdS well projection
which is much larger than the corresponding ZnS projection. Interestingly,
the size of the gap is governed essentially by the ZnS properties.
We also look at the radial
distribution of charge corresponding to the eigenstates. 
We can see the numerical values of the probability of presence
of electron and hole in the LUMO and HOMO in Table \ref{electron_size} 
 for several cases. The idea behind
the comparison is to investigate the role of the successively added
CdS layers. Both the electron and hole show an enhancement of their
presence in the well region; the effect is more pronounced for the
hole. 
When two CdS layers are added the hole is essentially trapped in the
well. In Table \ref{gaps} we show the
two highest valence (HOMO) and two lowest conduction (LUMO) levels.
The ZnS single-particle
spectrum presents an intrinsic degeneracy which is the consequence
of the absence of spin-orbit coupling: When a level is $\Gamma _{7}-\Gamma _{8}$
degenerate its orbital symmetry is $\Gamma _{5}$, because the spin-$1/2$
representation is $\Gamma _{6}$ and $\Gamma _{5}\times \Gamma _{6}=\Gamma _{7}+\Gamma _{8}$.
Similarly, the degeneracy $\Gamma _{6}-\Gamma _{8}$ indicates that
the underlying orbital symmetry is $\Gamma _{4}$.  

The excitonic spectrum is calculated by diagonalizing the exciton Hamiltonian in Eq.~(1) in
the configuration-interaction
method with as many valence and conduction states as necessary to
reach numerical convergence.
The relative intensities of optical transitions are next computed
 by following the procedure described in Ref. \cite{pb01}.  

In order to discuss the available experimental data, let us first
note that in QDQW structures like CdS/ZnS: in Ref.
\cite{ra98} the dispersion for dots with a mean diameter of $58$ \AA{}
is $\sigma =9$ \AA{}. On the other hand, in small nanocrystals,
theoretical results show big jumps in the energies when one adds just
one or two shells to a given NC: See Table \ref{gaps}. Our procedure
to fix the adequate size for a given set of experimental data \cite{le01}
is as follows: We first analyze several NC's of sizes differing by
one or two atomic shells around the experimental size and retain three such
structures yielding a good agreement with the experimental values
of size and optical absorption gap. The procedure is illustrated in 
Fig. \ref{absorption_zns}: We show the fine structure of absorption
for three different ZnS NC's of around $20$ \AA{} diameter. We can
see that the experimental onset from \cite{le01}, indicated by a
vertical dashed line, is in good agreement with our results if we
remember that the onset must have a finite width. Note that the successive
sizes in Fig. \ref{absorption_zns} correspond to the addition of
single atomic shell. In Fig. \ref{absorption_zns_2CdS} we show
the fine structure of a series of QDQW's based on a ZnS core of radius
$7.2$ \AA{}. The experimental values are given in the figure caption.
Finally, it is worth mentioning that the predicted photoluminescence
Stokes shift is enhanced by the presence of the barriers. In 
particular, the Stokes shift in the $R=14.8$ \AA{} CdS NC is $12$ meV, whereas the
$7.2/5.9/2.6$ QDQW, even slightly bigger in size, shows a shift of $53$ meV.

 In Fig. \ref{fig8} we  show a comparison between ZnS and CdS cores with ZnS/CdS and CdS/ZnS
  core-shell structures. The results are in accord with the experimental data of
  Ref. \cite{le01}: 
 The absorbance changes little between CdS and CdS/ZnS in the low-energy region, but the change is drastic
  between ZnS and ZnS/CdS. The former result has to do with a surface that is already fully passivated
  so that a thin barrier shell enhances the localization only slightly. On the other hand, in ZnS/CdS,
  the outer shell is a well and the spatial distributions are drastically modified, leading to a different
  absorption spectrum.

%\section{Conclusion}

\label{sec:Conclusion}To conclude, we have studied the electronic structure
and optical properties of spherical nanoheterostructures based on ZnS and CdS 
within a symmetry-based tight-binding approach. The
full single-particle spectrum, the low-energy excitonic states 
as well as
the relative intensities of optical transitions are calculated. 
A careful analysis of the spatial distribution of the DOS and the HOMO and LUMO
wavefunctions reveals that only two monolayers of CdS suffice to 
trap both the electron and the hole in the well. Our principal results concern
the excitonic fine structure yielding the absorption gap and the resonant 
photoluminescence Stokes shift. Their size and composition dependences
have been analyzed on a monolayer scale. We obtain a satisfactory
agreement with the available experimental values of the 
absorption onset without using any adjustable parameter. The predicted
Stokes shifts indicate a strong enhancement in the nanoheterostructures
as compared to simple nanocrystals.

This work has been supported by the Spanish Ministerio de Ciencia y Tecnolog{\'\i}a project MAT2002-00699.

\newpage

\begin{table}[ht]
\begin{tabular}{|c|c|c|c|}
\hline 
$r_{co}/r_{w}/r_{cl}$  &
 Core &
 Well &
 Clad \\
\hline
$7.2/3.2/0.0$&
 $0.592(0.323)$&
 $0.383(0.671$&
 $0.0(0,0)$\\
\hline
$7.2/5.9/0.0$&
 $0.322(0.118)$&
 $0.659(0.879)$&
 $0.0(0.0)$\\
\hline
$7.2/2.7/2.9$&
 $0.419(0.171)$&
 $0.328(0.590)$&
 $0.241(0.237)$\\
\hline
$7.2/5.9/2.6$&
 $0.211(0.046)$&
 $0.623(0.663)$&
 $0.158(0.289)$ \\
\hline
\end{tabular}

\protect

\caption{The probability of presence of electron(hole) in the LUMO(HOMO). The core radius
and the well and clad widths are given in the first column for each
case (in angstroms). }

\label{electron_size}
\end{table}

\begin{table}[hb]
\begin{tabular}{|c|c|c|c|c|}
\hline 
$r_{co}/r_{w}/r_{cl}$ (\AA) &
 H1 (eV) &
 H2 (eV) &
 L1 (eV)&
 L2 (eV) \\
\hline
$10.1/0.0/0.0$&
 $-0.322(8,7)$&
 $-0.378(8,6)$&
 $4.251(6)$&
 $4.598(7,8)$\\
\hline
$7.2/3.2/0.0$&
 $-0.156(8)$&
 $-0.167(6)$&
 $3.995(6)$&
 $4.291(8)$\\
\hline
$7.2/5.9/0.0$&
 $0.056(8)$&
 $0.036(6)$&
 $3.662(6)$&
 $3.922(8)$\\
\hline
$7.2/2.7/2.9$&
 $-0.046(8)$&
 $-0.058(6)$&
 $3.844(6)$&
 $4.112(8)$\\
\hline
$7.2/5.9/2.6$&
 $0.188(8)$&
 $0.166(6)$&
 $3.574(6)$&
 $3.809(8)$ \\
\hline
$0.0/14.8/0.0$&
 $-0.247(8)$&
 $-0.256(8)$&
 $2.988(6)$&
 $3.335(8)$\\
\hline
\end{tabular}

\protect

\caption{ZnS/CdS/ZnS: Energies of the two highest (lowest) occupied (unoccupied)
molecular orbitals, denoted by H1, H2 (L1, L2). The symmetry index
$n$ designating $\Gamma _{n}$ is shown in the parenthesis. The size
and composition are given in the first column. }

\label{gaps}
\end{table}

\begin{figure*}[p]
\caption{\label{Density-of-States:a}Density of States for three
NC's. Top: DOS in a ZnS QD of radius $R=10.1$ \AA. Middle(bottom):
A ZnS/CdS nanocrystal with a core of $R=7.2$ \AA{} and one (two)
CdS monolayer (s). In the middle and bottom figures the shaded region
corresponds to the DOS projected onto ZnS core region and the solid
line to the DOS projected onto the CdS well region.}
\end{figure*}

\begin{figure*}[p]

\caption{\label{absorption_zns}Excitonic fine structure 
in ZnS QD's. The experimental absorption onset is indicated by the
vertical dashed line. We show three similar sized QD's of around $10$ \AA{}
in radius.}
\end{figure*}

\begin{figure*}[p]
\caption{\label{absorption_zns_2CdS}Excitonic fine structure
in QDQW's. The core radii and shell thickness are all given in \AA. 
 The respective experimental absorption onsets are: (a) $3.85$ eV,
(b) $3.55$ eV, (c) $3.76$ eV, (d) $3.45$eV.}
\end{figure*}

\begin{figure*}[p]
\caption{\label{fig8}Excitonic fine structure
in core and core-shell nanostructures. 
The core radii and shell thickness are all given in \AA.} 

\end{figure*}

\includegraphics[  angle=270, width=1.0\columnwidth, origin=c, keepaspectratio]{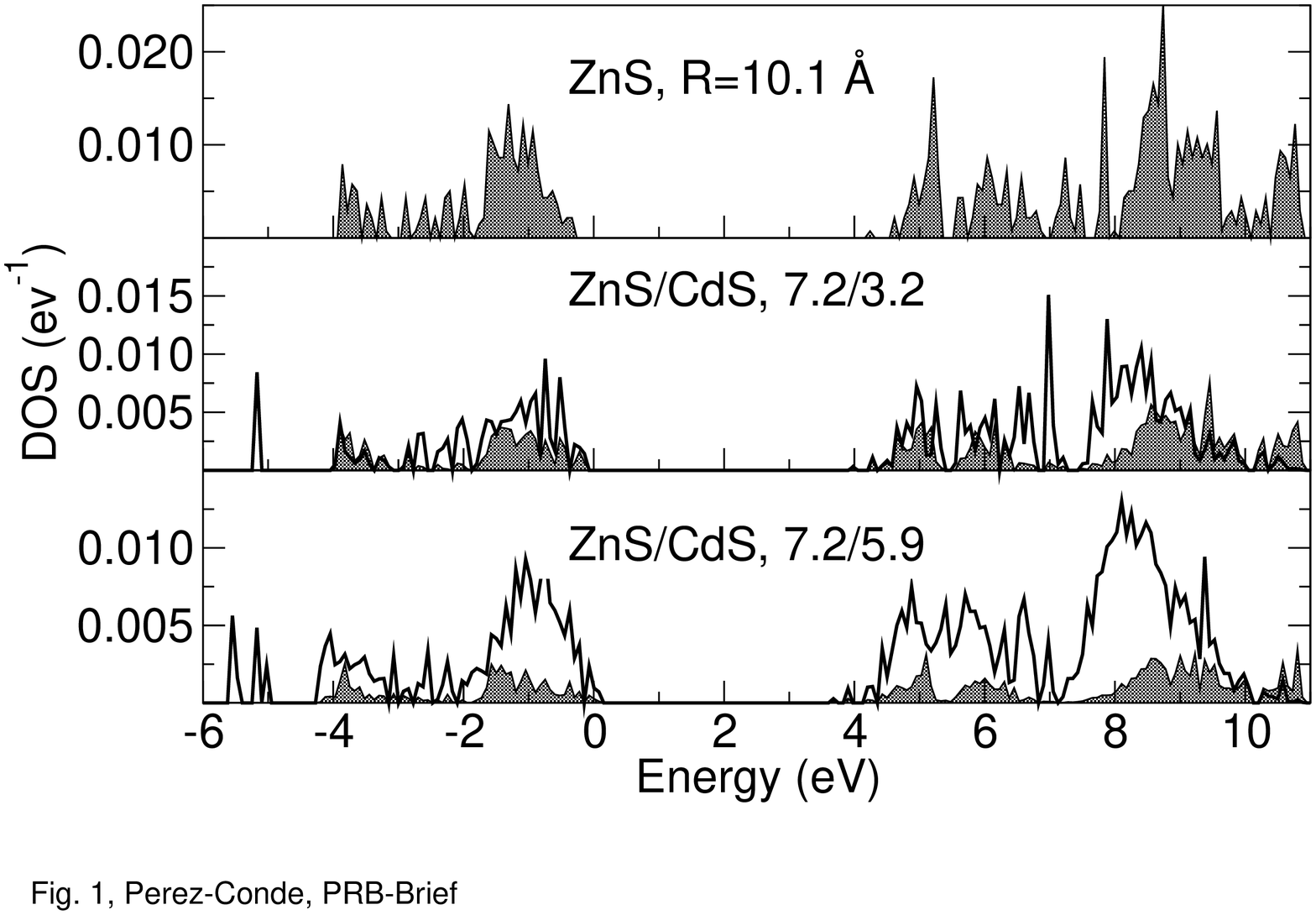}

\includegraphics[  angle=270, width=1.0\columnwidth, origin=c,  keepaspectratio]{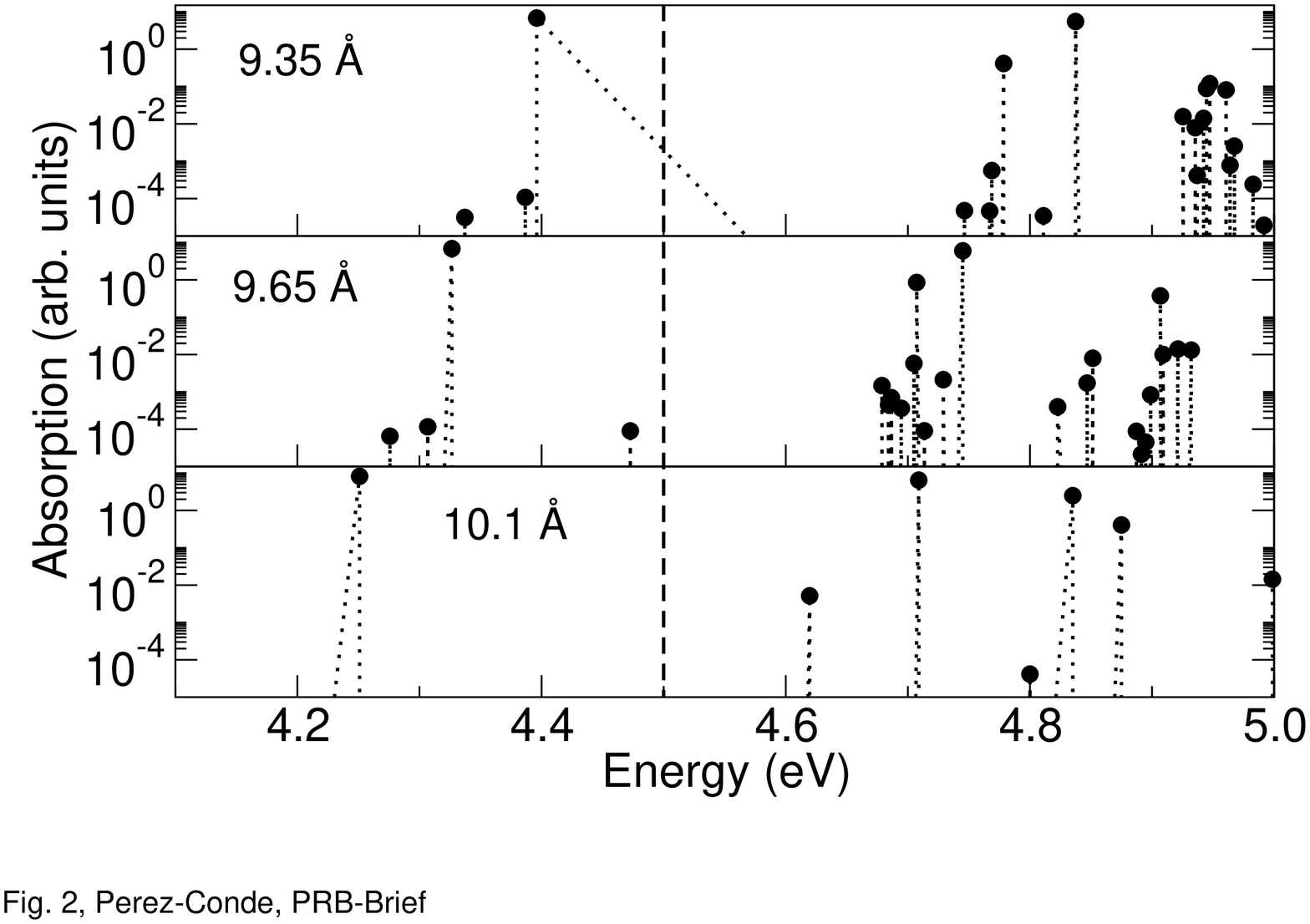}

\includegraphics[  angle=270, width=1.0\columnwidth, origin=c,  keepaspectratio]{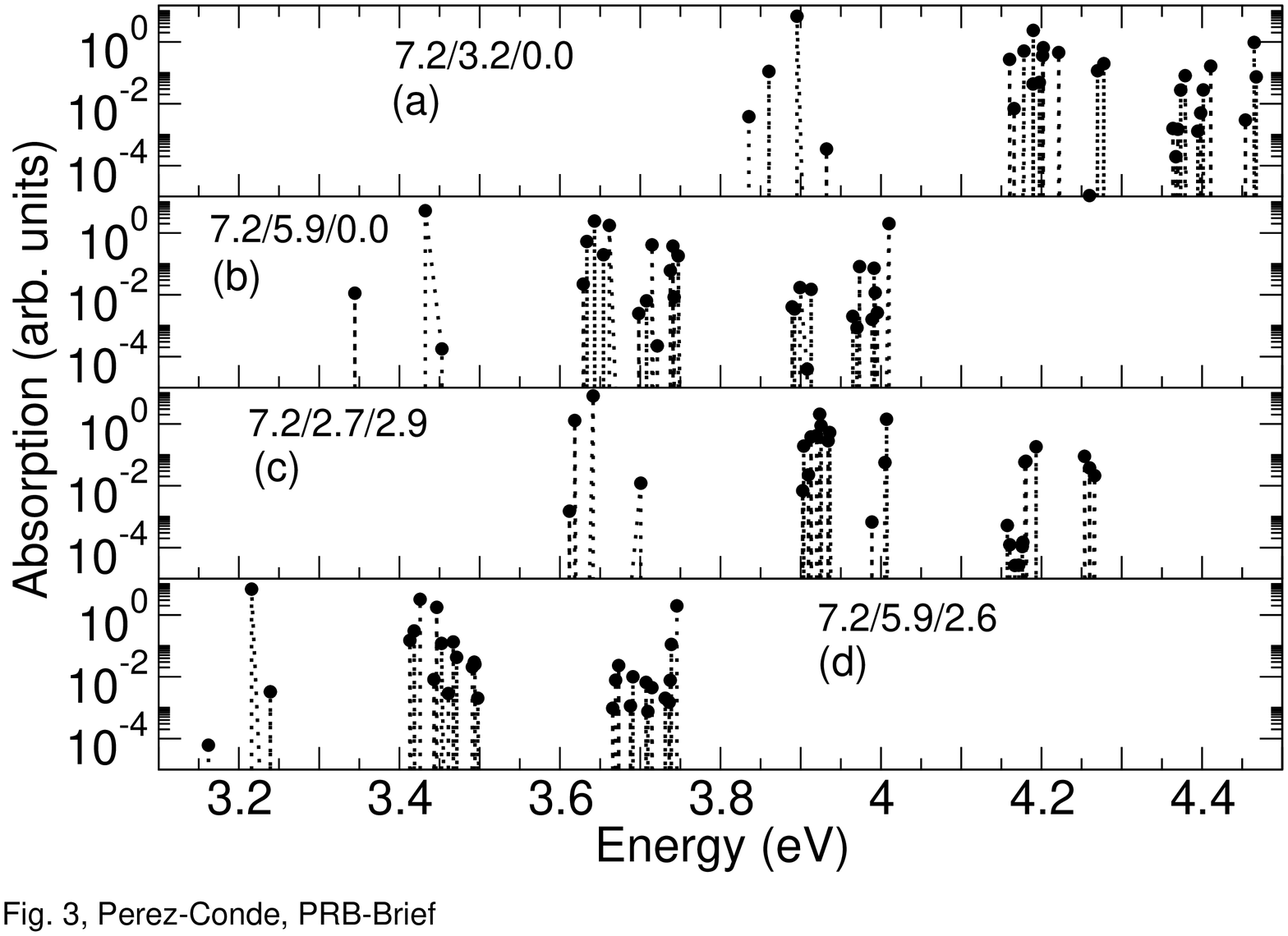}

\includegraphics[  angle=270, width=1.0\columnwidth, origin=c,  keepaspectratio]{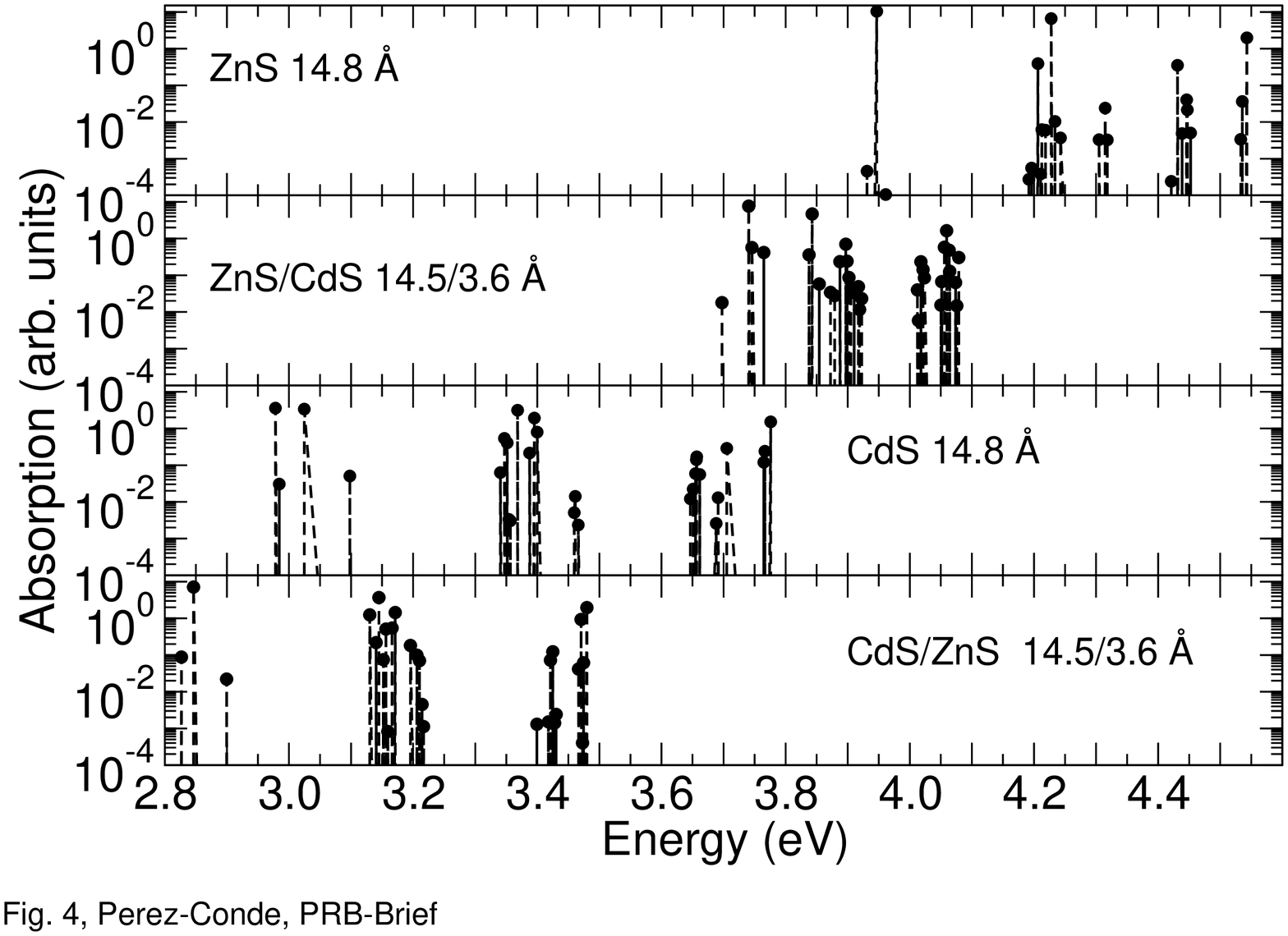}


\begin{thebibliography}{10}
\bibitem{mk96}A.~Mews, A.~V.~Kadavanich, U.~Banin, and A.~P.~Alivisatos, Phys.
Rev. B \textbf{53}, R13 242 (1996). 
\bibitem{sm94}D.~Schooss, A.~Mews, A.~Eychm\"{u}ller, and H.~Weller, Phys. Rev.
B \textbf{49}, 17 072 (1994). 
\bibitem{km99}F.~Koberling, A.~Mews, and T.~Basch\'{e}, Phys. Rev. B \textbf{60},
1921 (1999). 
\bibitem{le01}R. B. Little, M. A. El-Sayed, G. W. Bryant, and S. Burke, J. Chem.
Phys. \textbf{114}, 1813 (2001). 
\bibitem{ra98}C. Ricolleau, L. Audinet, M. Gandais, and T. Gacoin, Thin Solid Films,
\textbf{336}, 213 (1998). 
\bibitem{dr97}B. O. Dabbousi, J. Rodriguez-Viejo, F. V. Mikulec, J. R. Heine, H. Mattoussi, R. Ober, K. F. Jensen, and M. G. Bawendi
J. Phys. Chem.B  \textbf{101}, 9463 (1997). 
\bibitem{hs99}E. Hao, H. P. Sun, Z. Zhou, J. Liu, B. Yang, J. C. Shen, Chem. Mater. \textbf{11}, 3096 (1999). 
\bibitem{jb98}W.~Jask\'{o}lski and G.~W.~Bryant, Phys. Rev. B \textbf{57}, R4237
(1998). 
\bibitem{pb02}J. P\'{e}rez-Conde and A. K. Bhattacharjee, phys. stat. sol. (b)
\textbf{229}, 485 (2002). 
\bibitem{pb01}J. P\'{e}rez-Conde and A. K. Bhattacharjee, Phys. Rev. B \textbf{63},
245318 (2001). 
\bibitem{pb01b}J. P\'{e}rez-Conde, A. K. Bhattacharjee, M. Chamarro, P. Lavallard, V. D. Petrikov, and A. A. Lipovskii,
 Phys. Rev. B \textbf{64}, 113303 (2001). 
\bibitem{pb99}J. P\'{e}rez-Conde and A. K. Bhattacharjee, Solid State Comm. \textbf{110},
259 (1999). 
\bibitem{bj01}G.~W.~Bryant and W.~Jask\'{o}lski, phys. stat. sol. (b) \textbf{224},
751 (2001). 
\bibitem{xb02}R.-H. Xie, G. W. Bryant, S. Lee, and W. Jaskolski, Phys. Rev. B \textbf{65},
235306 (2002). 
\bibitem{vh83}P. Vogl, H. P. Hjalmarson, and J. D. Dow, J. Phys. Chem. Solids \textbf{44},
365 (1983). 
\bibitem{ks82}A.~Kobayashi, O.~F.~Sankey, and J.~D.~Dow, Phys. Rev. B \textbf{25},
6367 (1982).
\bibitem{ll89}P.~E.~Lippens and M.~Lannoo, Phys. Rev. B \textbf{39}, 10935 (1989).  
\bibitem{be97}D. Bertho (unpublished), cited by V. Albe, Doctoral thesis, Universit\'{e}
Montpellier II (1997). 
\bibitem{kd63}G. F. Koster, J. O. Dimmock, R. G. Wheeler and H. Statz, \textit{Properties
of the Thirty-Two Point Groups (MIT Press, Cambridge, MA, 1963)}. 
\bibitem{lp98}K. Leung, S. Pokrant, and K. B. Whaley, Phys. Rev. B \textbf{57},
12 291 (1998). 
\bibitem{wz96} L. W. Wang and A. Zunger, Phys. Rev. B \textbf{53}, 9579 (1996).    
\bibitem{bk02} T. B. Boykin, G. Klimeck, R. C. Bowen, and F. Oyafuso,
Phys. Rev. B \textbf{66}, 125207 (2002).
%\bibitem{mn93}C. B. Murray, D. J. Norris, and M. G. Bawendi, J. Am. Chem. Soc. \textbf{115},
%8706 (1993). 
\end{thebibliography}
\end{document}